\documentclass[iop]{emulateapj}

\shorttitle{Quasar Host Galaxies on SDSS Stripe 82}
\shortauthors{Matsuoka et al.}

\usepackage{color}

\begin{document}


\title{Massive Star-Forming Host Galaxies of Quasars on Sloan Digital Sky Survey Stripe 82}


\author{Yoshiki Matsuoka\altaffilmark{1,2}, Michael A. Strauss\altaffilmark{1}, Ted N. Price III\altaffilmark{1,3}, and Matthew S. DiDonato\altaffilmark{1,4}}


\altaffiltext{1}{Princeton University Observatory, Peyton Hall, Princeton, NJ 08544, USA}
\altaffiltext{2}{National Astronomical Observatory of Japan, Mitaka, Tokyo 181-8588, Japan}
\altaffiltext{3}{Eastdil Secured, 40 West 57$^{\rm th}$ Street, New York, NY 10019, USA}
\altaffiltext{4}{Haddonfield Memorial High School, Haddonfield, NJ 08033, USA}


\begin{abstract}
The stellar properties of about 800 galaxies hosting optically luminous, unobscured quasars at $z < 0.6$ are analyzed.
Deep co-added Sloan Digital Sky Survey (SDSS) images of the quasars on Stripe 82 are decomposed into nucleus and host galaxy
using point spread function and S\'{e}rsic models.universe
The systematic errors in the measured galaxy absolute magnitudes and colors are estimated to be less than 0.5 mag and 0.1 mag,
respectively, with simulated quasar images.
The effect of quasar light scattered by the interstellar medium is also carefully addressed.
The measured quasar-to-galaxy ratio in total flux decreases toward longer wavelengths, from $\sim$8 in the $u$ band to $\sim$1
in the $i$ and $z$ bands.
We find that the SDSS quasars are hosted exclusively by massive galaxies (stellar mass $M_{\rm star} > 10^{10} M_{\odot}$), 
which is consistent with previous results for less luminous narrow-line (obscured) active galactic nuclei (AGNs).
The quasar hosts are very blue and almost absent on the red sequence, showing stark contrast to the color--magnitude distribution of normal galaxies.
The fact that more powerful AGNs reside in galaxies with higher star-formation efficiency may indicate that negative AGN feedback, if it exists, is 
not concurrent with the most luminous phase of AGNs.
We also find positive correlation between the mass of supermassive black holes (SMBHs; $M_{\rm BH}$) and host stellar mass, but the 
$M_{\rm BH} - M_{\rm star}$ relation is offset toward large $M_{\rm BH}$ or small $M_{\rm star}$ compared to the local relation.
While this could indicate that SMBHs grow earlier than do their host galaxies, such an argument is not conclusive,
as the effect may be dominated by observational biases.
\end{abstract}





\section{Introduction}

The interrelation between supermassive black holes (SMBHs) and their host galaxies is a central issue of astrophysics today.
The unexpectedly tight correlation
between the mass of SMBHs ($M_{\rm BH}$) and the velocity dispersion ($\sigma_*$) or mass of their host bulges 
\citep[e.g., ][]{dressler89,kormendy93,magorrian98,merritt01,mclure02,haring04,gultekin09}
suggests that the two coevolve, or at least strongly influence each other.
The observed scatter in the $M_{\rm BH} - \sigma_*$ relation is explained almost solely by
measurement errors \citep{ferrarese00,gebhardt00}.
Such a tight correlation suggests a fine-tuning process, perhaps a combination of internal physics within galaxies or dark halos,
and mass averaging via successive mergers as predicted in hierarchical galaxy formation models
\citep[see the recent review by ][and references therein]{kormendy13}.
Among the possible physical processes that can drive the co-evolution of SMBHs and host galaxies, the most compelling today
is negative feedback caused by active galactic nuclei \citep[AGNs; e.g.,][]{ciotti07,ostriker10,fabian12}.
It has been suggested that the energy input from AGNs, which may have been triggered along with active starbursts by 
interactions/mergers of gas-rich galaxies \citep[e.g.,][]{barnes91}, 
could quench star formation in the host galaxies by expelling the cold gas
from galaxies \citep[quasar-mode feedback; e.g.,][]{dimatteo05,springel05} and/or heating the gas in dark halos 
\citep[radio-mode feedback; e.g.,][and references therein]{mcnamara07}.

The AGN feedback model is gaining broad support in part because it may give a solution to longstanding problems in
standard galaxy formation models \citep[e.g., ][]{somerville08}; 
for example, it is currently the most compelling process to reconcile the very different shapes of the dark halo mass function predicted 
in the $\Lambda$CDM cosmology and the observed galaxy stellar mass function, especially at the high mass end 
\citep[e.g.,][]{bower06,croton06,dave11,choi11}.
In addition, recent observations of AGN-driven gas outflows \citep[e.g., ][]{nesvadba06,feruglio10,greene11,cano-diaz12} and
extended emission-line regions photoionized by AGNs \citep[e.g.,][]{fu09,husemann10,greene11,matsuoka12,keel12,liu13a,liu13b} 
strongly suggest that AGNs have significant impact on their host galaxies.
However, all these arguments are only circumstantial evidence for the putative AGN feedback.
Stellar populations of AGN hosts are expected to provide a key piece of information, since the end effect of AGN feedback
should appear in stellar properties.

One of the earliest studies of the stellar populations of quasar hosts was presented by \citet{boroson82}
who found that the off-nucleus spectrum of the quasar 3C 48 was dominated by hot young stars.
\citet{kotilainen94} analyzed imaging data of nearby Seyfert 1s and found that 
near-infrared colors of their hosts are similar to those of moderately luminous starburst galaxies and that
host luminosity is positively correlated with AGN luminosity.
\citet{ronnback96}, studying radio-loud and radio-quiet quasars at $0.4 < z < 0.8$, discovered that the quasars 
reside in luminous galaxies whose optical colors are consistent with late-type spirals and irregular galaxies.
More recent studies have benefited greatly from the advent of {\it Hubble Space Telescope} ({\it HST}), since resolution
of the host galaxies from extremely bright nucleus is the key to accurate quantification of host properties.
\citet{bahcall97} showed that luminous quasars reside preferentially in luminous galaxies with diverse morphology.
The excess detection of close companions, tidal features, and dense surrounding environments led them to 
conclude that gravitational interactions are important in triggering quasar activity.
In addition, \citet{kirhakos99} found that quasar hosts are 0.5--1.0 mag bluer in $V - I$ than normal galaxies
-- a sign of active star formation connected to quasars.
On the other hand, \citet{mclure99} claimed that quasars are almost exclusively hosted by massive ellipticals
(the mean half-light radius was found to be $\sim$10 kpc) with $R - K$ colors consistent with old stellar populations.
At least part of the discrepancy between the Bahcall et al. and McLure et al. results may be attributed to
the different filter choices 
and different modeling methods.
At higher redshifts up to $z \sim 3$, \citet{jahnke04} showed that blue ultraviolet colors of host galaxies indicate
recent starburst or ongoing star formation activity, which was later reconfirmed using spatially decomposed spectra
by \citet{jahnke07}.
A similar result in terms of enhanced blue light in quasar host galaxies was also obtained by \citet{sanchez04}.

Recently, the Sloan Digital Sky Survey \citep[SDSS;][]{york00} has revolutionized this field via detection
of large samples of narrow-line AGNs with low \citep[e.g., ][]{kauffmann03,hao05} and high \citep[e.g.,][]{zakamska03, reyes08} luminosity.
In these obscured populations, the host galaxies are much more easily observed than in unobscured objects 
\citep[e.g.,][]{zakamska06,liu09}.
Obscured AGNs have also been investigated with X-ray observations at redshifts up 
to $z = 1$ and above \citep[e.g.,][]{nandra07, georgakakis08, silverman08}.
Overall, the above studies are consistent in finding that obscured AGNs reside preferentially in massive galaxies whose
colors are systematically bluer than those of normal galaxies with similar stellar mass.
Of particular note is the highest AGN fraction in the green valley of the color-magnitude diagram (CMD) as claimed 
by some authors \citep[e.g., ][]{schawinski10}, which is consistent with (but not necessarily indicative of) a scenario 
in which AGNs are responsible for transforming blue star-forming galaxies into red quiescent galaxies.
AGNs found in post-starburst galaxies \citep[e.g.,][]{brotherton99,brotherton02} may represent this stage.
There are also some indications that higher-luminosity AGNs reside in bluer galaxies \citep[e.g.,][]{kauffmann03, salim07},
which seems to contradict the idea that negative AGN feedback on star formation is concurrent with the period of highest nuclear activity.
At the same time, detailed analyses of obscured quasars have revealed the presence of significant amount of scattered quasar light
in host galaxies \citep{zakamska06,liu09}.
If not handled properly, this could result in a serious bias in which the hosts of more luminous AGNs are observed to be bluer, 
since the AGN spectral energy distribution is usually much bluer than that of starlight.

Despite the above intense efforts, 
one of the biggest pieces of the puzzle still missing is the general nature of galaxies hosting ``classical" quasars, i.e., hosts of optically-luminous unobscured quasars.
Since different types of AGN may have different hosts and may play different roles in galaxy evolution \citep[e.g.,][]{hickox09}, 
it is critically important to study the hosts of all types.
The simplest AGN unification model \citep[see the review article of][]{antonucci93} assumes that unobscured (type-1) and obscured (type-2) AGNs 
are intrinsically the same except for the orientation of the obscuring material relative to the observer's line-of-sight, 
but the actual situation is likely to be more complicated.
For example, obscured AGNs are more likely to be embedded in dusty circumnuclear environments than are unobscured ones \citep{elitzur12}.
In the merger-driven evolution scenario of galaxies and SMBHs described by \citet{hopkins06}, the dusty starburst/AGN phase (sometimes observed as
ultra-luminous infrared galaxies, ULIRGs) precedes the dust-free phase observed as optically luminous quasars \citep[see also][]{sanders88}.
If this is the case, then obscured and unobscured AGNs would represent different stages of galaxy evolution.

In this work, we present a statistical analysis of host galaxies of luminous unobscured quasars at low redshifts ($z < 0.6$). 
By making use of the deep co-added SDSS images on Stripe 82 on the celestial equator, we successfully decompose host galaxies from
bright quasar nuclei and investigate the stellar properties of more than 800 sources.
This paper is organized as follows. 
The data and sample are presented in Section 2.
Section 3 describes the method of analysis we use to decompose the source images into quasar nuclei and host galaxies.
The main results appear in Section 4.
Then, in Section 5, we discuss the implication of the results as well as the possible drawbacks of the present analysis, 
including quasar light contamination in measured host brightness.
A summary follows in Section 6.
The cosmological parameters of $H_0 = 70$ km s$^{-1}$, $\Omega_{\rm M} = 0.3$, and $\Omega_{\rm \Lambda} = 0.7$
are assumed throughout the paper.
All the magnitudes are presented on the AB system.

\section{Data and Sample Selection \label{sec:data}}


The data and catalogs used in this work were obtained through a set of SDSS projects; SDSS-I \citep{york00}, the SDSS-II 
Supernova Survey \citep{frieman08}, and the SDSS-III Baryon Oscillation Spectroscopic Survey \citep[BOSS;][]{dawson13}. 
SDSS uses a dedicated 2.5 m telescope \citep{gunn06} at Apache Point Observatory, New Mexico.
The telescope is equipped with a mosaic CCD camera \citep{gunn98} and twin multi-object fiber spectrographs developed 
for SDSS-I and upgraded for BOSS \citep{smee12}.
We make use of the deep stacked images of Stripe 82, the 2.5$^\circ$-wide (in declination) stripe on the celestial equator
from about 20$^{\rm h}$ to 6$^{\rm h}$ (in right ascension).
Those images have been created by co-adding multiple scans of the stripe obtained through SDSS-I and the Supernova Survey
\citep{annis11}.
Typically about 20 scans contribute to each piece of the sky, giving an effective exposure time of $\sim$1000 s.
The 50\% completeness limits for point sources are (23.6, 24.6, 24.2, 23.7, 22.3) mag in the ($u$, $g$, $r$, $i$, $z$) bands, which are 
approximately 2 mag deeper than a single SDSS exposure.
The mean full widths at half maximum of the point spread functions (PSFs) are $\sim$1\arcsec.3 in the $u$, $g$ bands and $\sim$1\arcsec.1 in
the $r$, $i$, and $z$ bands.

\begin{figure}
\epsscale{1.0}
\plotone{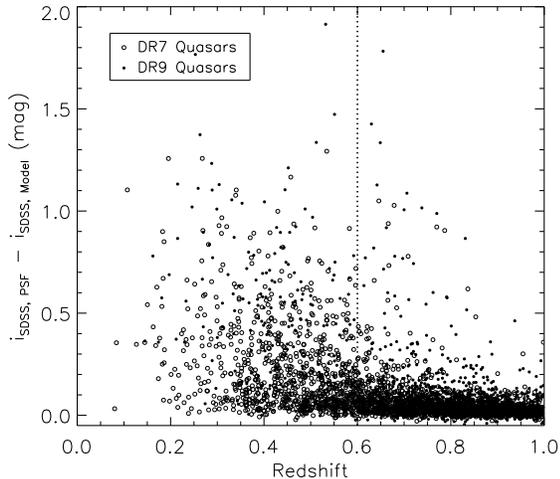}
\caption{Difference between the two sets of SDSS $i$-band magnitudes, $i_{\rm SDSS, PSF}$ (PSF magnitudes) and 
$i_{\rm SDSS, Model}$ (model magnitudes), of the quasars on Stripe 82 as a function of redshift.
The open and filled circles represent the DR7 and DR9 quasars, respectively. 
The dotted line shows the redshift limit ($z = 0.6$) applied in this work.}
\label{fig:checkresolved}
\end{figure}

\begin{figure}
\epsscale{1.0}
\plotone{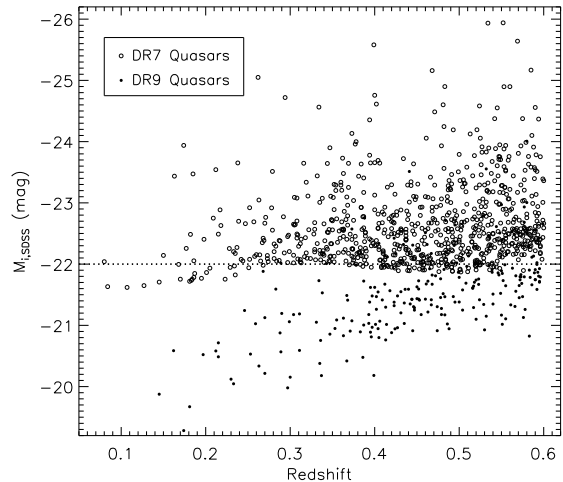}
\caption{Redshifts and absolute magnitudes $M_{i, {\rm SDSS}}$ of the quasar sample
(open circles: DR7 quasars, filled circles: DR9 quasars).
The dotted line represents the luminosity threshold of the original DR7 quasar selection \citep{schneider10}.
}
\label{fig:sample}
\end{figure}

The sample is selected from the fifth SDSS quasar catalog \citep{schneider10} based on data release (DR) 7 
\citep{abazajian09} and the first BOSS quasar catalog \citep{paris12} based on DR9 \citep{ahn12}.
The former catalog contains quasars with $M_i < -22$ mag, while the latter contains fainter objects down to
$M_i (z = 2) = -20.5$ mag \citep[corresponding to $M_i = M_i (z = 0) \simeq -19.9$ mag;][]{richards06}.
In order to have a rough estimate of the resolved fraction as a function of redshift, we check the difference between the two sets of SDSS magnitudes, 
the PSF magnitudes ($m_{\rm SDSS, PSF}$) and the model magnitudes ($m_{\rm SDSS, Model}$), of those quasars on Stripe 82.
Since $m_{\rm SDSS, PSF}$ and $m_{\rm SDSS, Model}$ are designed to measure the total brightness of unresolved and resolved sources, respectively,
the difference between the two is larger for more extended sources.
Since a good fraction of the quasars seem to be resolved up to $z \sim 0.6$ as plotted in Figure \ref{fig:checkresolved}, 
we decided to apply the present analysis to the objects at $z < 0.6$.
After removing the objects with nearby bright sources (the rejection criteria are detailed in the next section), we have 1041 (802 DR7 and 239 DR9) 
quasars which constitute our initial sample.
Those objects have a variety of SDSS target flags.
Most of them have been targeted as quasar or galaxy targets 
for various reasons, based on colors, extendedness, variability, and/or detection at other wavelengths.
The total number of objects targeted as quasar candidates is roughly equal to those targeted as galaxies in our sample.
We show the distribution of redshifts and absolute magnitudes $M_{i, {\rm SDSS}}$\footnote{
Hereafter $i_{\rm SDSS}$ denotes the SDSS PSF magnitude ($= i_{\rm SDSS, PSF}$) taken from the original quasar catalogs
and corrected for Galactic extinction from \citet{schlegel98}.
$M_{i, {\rm SDSS}}$ is the absolute magnitude calculated from $i_{\rm SDSS}$ with a $k$-correction to $z = 0$.
}
of the initial sample in Figure \ref{fig:sample}.
The $k$-correction has been applied following \citet{richards06}.
Most of the objects are at $z > 0.3$; the median redshift is $\langle z \rangle = 0.47$.
Note that some of the DR7 quasars fall below the luminosity threshold originally applied by \citet{schneider10}
because of the slightly different $k$-corrections applied.
We could select more low-luminous AGNs at low redshifts from the SDSS main galaxy sample \citep{strauss02}, but we defer this to future work.

We additionally retrieve the nuclear properties ([\ion{O}{3}] $\lambda5007$ line luminosity, SMBH mass, and Eddington ratio) 
of the DR7 quasars from \citet{shen11}.
Most of the properties were estimated based on the quasar emission lines contained in the SDSS spectral data.
All the SMBH masses used in this work were derived from the single-epoch measurement of H$\beta$, which is believed to 
be more reliable than any other known single-epoch estimators \citep{shen12,shen13}.
The following results and arguments related to those nuclear properties include only the DR7 quasars.

\section{Image Analysis \label{sec:analysis}}

\subsection{Measurements of Host Galaxy Brightness \label{subsec:imageanalysis}}

\begin{figure}
\epsscale{1.0}
\plotone{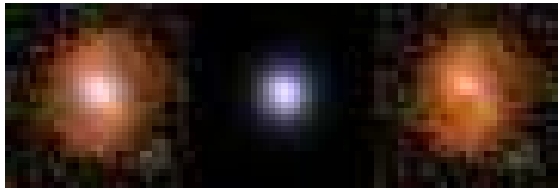}
\caption{Three-color composite images of SDSS $J$002450.50$+$003447.7 at $z = 0.524$. 
The SDSS $i$, $r$, $g$-band images were used to create the red--green--blue composites with the recipe of \citet{lupton04}.
Left: the original image, middle: the best-fit PSF model, right: the PSF-subtracted image.
The same flux scaling is applied for the three panels so that the brightness and colors can be directly compared by eye.
The size of each panel corresponds to 100 kpc $\times$ 100 kpc at the object's distance.
}
\label{fig:snapshot}
\end{figure}

We work directly on co-added SDSS images created by \citet{annis11}.
We show three-color composite images of one of the quasars in Figure \ref{fig:snapshot} as an illustration.
The analysis is carried out separately for images of the same object taken through different filters.
In order to avoid any additional complexity and/or systematic uncertainty arising from source deblending, we reject those objects
whose images are possibly affected by nearby sources.
If a bright source ($< 15$ mag in any of the $u$, $g$, $r$, $i$, and $z$ bands) is present in the 100 pixel $\times$ 100 pixel 
(40\arcsec $\times$ 40\arcsec) image 
cutout centered on a quasar, then the quasar is thrown away.
For a fainter nearby source of magnitude $m$ found in the image cutout, all the pixels within $10.0 - 0.5 (m - 15.0)$ pixels
of the source (empirically determined to be large enough to avoid any significant light contamination)
are excluded from the fitting procedure described below.
We check the fraction of the rejected pixels around each quasar, and if the fraction exceeds 0.5 in any one of the five annuli centered
on the quasar with inner/outer radii of $n$/$n+1$ pixels ($n = 0, 1, 2, 3, 4$), then the quasar is excluded from the sample.
The above criteria reduce the number of the quasars by about 6 \% and result in our initial sample of 1041 objects.
Although sky subtraction has been already done for the co-added images, we found that the background values around
some of the quasars have positive or negative offset levels.
This may be due to residual sky subtraction issues in SDSS photometric processing, as described in \citet{abazajian09}.
We measure this local offset by iterative $\sigma$-clipping within the 100 pixel $\times$ 100 pixel image cutout, and correct for it.


We fit the azimuthal-average radial profile of each object, $I_{\rm obj} (R)$, with an optimized combination of the PSF 
and the \citet{sersic68} function \citep[a concise reference to the S\'{e}rsic function is found in][]{graham05}.
We do not use a more complicated procedure like two-dimensional fitting in this study, partly due to the image quality 
of SDSS (which is significantly worse than those of {\it HST} observations), and partly because
we are interested only in total brightness (not in the exact values of structural parameters).
The pixel coordinates of the quasar centroid are obtained from the celestial coordinates listed in the original quasar catalogs
and the astrometric calibration stored in the header of the co-added images (\texttt{fpC} frames).
The PSF at the position of each source in each band is extracted from the relevant SDSS measurements
(contained in the \texttt{psField} products); in the SDSS data processing,
the PSF in each CCD frame was determined from 15--20 bright stars using the Karhunen--Lo\'{e}ve transform technique \citep{lupton01}, 
which was then synthesized into the PSF of a co-added image with suitable weighting \citep{annis11}.
The amplitude $I_{\rm PSF}$ is the only adjustable parameter associated with the central point source in our analysis.
The S\'{e}rsic function has the form
\begin{equation}
I (R) = I_{\rm e}\ {\rm exp} \left(-b_n \left[ \left( \frac{R}{R_{\rm e}} \right) ^\frac{1}{n} -1 \right]   \right) ,
\end{equation}
where $R_{\rm e}$ is the effective radius enclosing half of the total light and $I_{\rm e}$ is the intensity at 
the radius $R = R_{\rm e}$.
The S\'{e}rsic index $n$ describes the overall profile shape.
The constant $b_n$ is uniquely determined once $R_{\rm e}$ and $n$ are given, so that $R_{\rm e}$ meets its definition
as the half-light radius.
Therefore the fitting procedure involves four free parameters of two profile components; $I_{\rm PSF}$ (PSF) and
$I_{\rm e}$, $n$, $R_{\rm e}$ (the S\'{e}rsic function).
The S\'{e}rsic function is convolved with the corresponding PSF to accommodate the effect of seeing.
The object's radial profile $I_{\rm obj} (R)$ is measured in bins one pixel wide, i.e., in annuli with the inner/outer 
radii of $n$/$n+1$ ($n = 0, 1, 2, ...$) pixels centered on the object, out to the pixel scale corresponding to 30 kpc
at the object's distance.
The associated error is estimated from the root mean square (rms) of the data values in each bin.
Any bins in which the mean flux is less than three times the associated error are excluded from the analysis.
We do not perform the analysis when there are less than five profile points available.
We keep track of all the discarded objects in each band throughout the fitting procedure for later discussion.


\begin{figure}
\epsscale{1.0}
\plotone{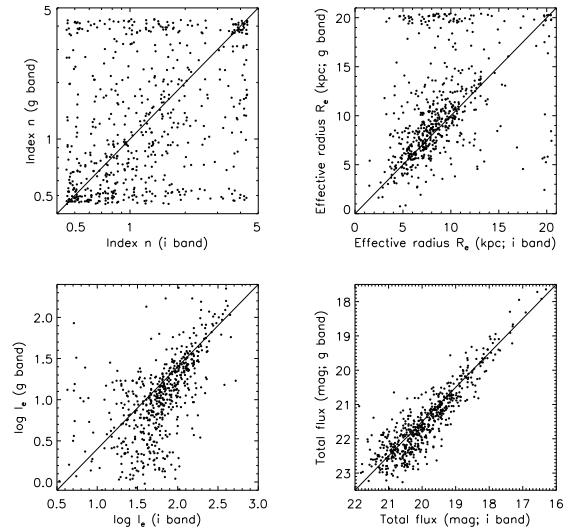}
\caption{Cross-comparison between the $g$ and $i$ bands of the best-fit parameters $n$ (top left), $R_{\rm e}$ (top right), 
log $I_{\rm e}$ (bottom left)
of the S\'{e}rsic profile as well as the integrated flux out to $R = 30$ kpc ($F_{\rm Galaxy}$; bottom right).
Small random offsets have been applied to improve visibility.
The typical fitting uncertainty is $\sigma_{{\rm log} n} = 0.25$, $\sigma_{R_{\rm e}} < 1.0$ kpc (i.e., less than the grid interval),
$\sigma_{{\rm log} I_{\rm e}} = 0.13$, and $< 0.1$ mag for the integrated flux.
The solid lines in the top two panels represent the locus where the results of the two bands are identical, 
while those in the bottom panels represent the approximate mean color of $g - i = 1.5$.
}
\label{fig:stat_initfit}
\end{figure}

In order to minimize parameter degeneracy and/or physically unreasonable regions of parameter space,
we carry out the fitting analysis in two steps as follows.
The first step is conducted in the $i$ band to determine the values of $n$ and $R_{\rm e}$ of each galaxy.
The choice of $i$ band is made since the quasar-to-galaxy contrast is in general lower at longer wavelength 
(i.e., galaxies are redder than quasars), making the galaxy component more visible, and
the SDSS signal-to-noise ratio is much better in the $g$, $r$, and $i$ bands than in the other two bands.
Also, the seeing tends to be better at longer wavelengths.
$I_{\rm PSF}$ is tentatively calculated by assuming that the source profile has negligible contribution from 
the S\'{e}rsic component at $R < 2$ pixel.
Then we subtract the PSF from the source and fit the residual profile at $R > 2$ pixel with the S\'{e}rsic function.
We discard those objects whose residual flux is less than 10 \% of the subtracted PSF flux.
The S\'{e}rsic amplitude $I_{\rm e}$ is varied freely, while $n$ and $R_{\rm e}$ are changed among values on a pre-fixed grid; 
$n$ can take ten values logarithmically stepped from 0.5 to 4.0, and $R_{\rm e}$ can take twenty values linearly stepped 
from 1 to 20 kpc.
We also conducted the above procedure in the $g$ band and cross-compared the results with those in the $i$ band, as shown in
Figure \ref{fig:stat_initfit}.
Clearly, the total fluxes contained in the S\'{e}rsic functions of the two bands correlate fairly well despite the presence of scatter
in intrinsic galaxy color, while none of the individual free parameters $n$, $R_{\rm e}$, or $I_{\rm e}$ shows better correlation.
This suggests that our approach works well to reproduce the overall flux of the non-PSF component, 
even if the individual parameters are not constrained very accurately.

\begin{figure}
\epsscale{1.0}
\plotone{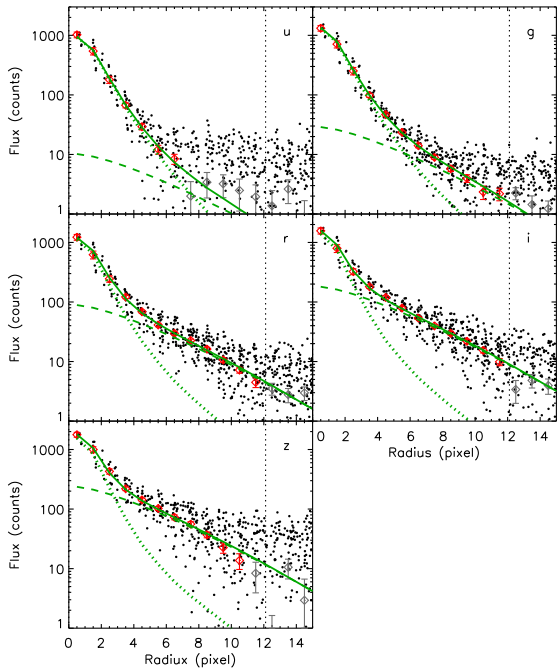}
\caption{Fitting results for the quasar SDSS $J$002450.50$+$003447.7 at $z = 0.524$ (the same object as shown in Figure \ref{fig:snapshot})
in the $u$ (top left), $g$ (top right), $r$ (middle left), $i$ (middle right), and $z$ (bottom) bands.
The dots correspond to individual CCD pixels around the quasar.
The diamonds represent the azimuthally averaged radial profile; those used in the fitting are in red, while those 
discarded for quality reasons (see text) are in gray.
The best-fit profiles are shown by the green lines; the PSF (dotted lines) and the S\'{e}rsic (dashed lines) components 
as well as their sum (solid lines).
The vertical dotted lines show the pixel scale corresponding to 30 kpc at the object's distance.
}
\label{fig:proffit}
\end{figure}

\begin{figure}
\epsscale{1.0}
\plotone{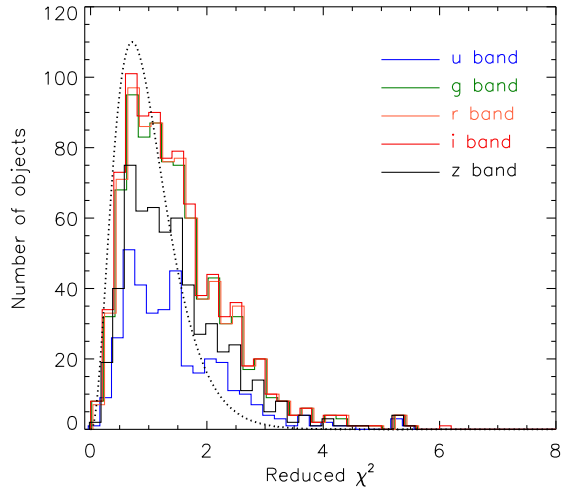}
\caption{Distributions of the reduced $\chi^2$ values of the best fits.
The blue, green, orange, red and black solid lines represent the $u$, $g$, $r$, $i$, and $z$ bands, respectively.
They are slightly offset from each other to improve visibility.
The dotted line represents the expected $\chi^2$ distribution (arbitrary scaled; the mean degree of freedom in the $i$ band is used for the calculation).
}
\label{fig:chi2}
\end{figure}

Second, we use the $n$ and $R_{\rm e}$ values determined above (in the $i$ band) and 
fit the combined PSF plus S\'{e}rsic function to the observed profile in each band by varying $I_{\rm PSF}$ and $I_{\rm e}$ simultaneously.
We show one of the good fitting results in Figure \ref{fig:proffit} to illustrate our procedure; this is the same object as shown in Figure \ref{fig:snapshot}.
The distribution of the reduced $\chi^2$ values of the best fits peaks around 1 and is fairly close to the expected $\chi^2$ distribution in all the five bands, 
as shown in Figure \ref{fig:chi2}, suggesting that our model reproduces the observed source profiles reasonably well.
We measure the galaxy flux ($F_{\rm Galaxy}$) by integrating the best-fit S\'{e}rsic function out to $R = 30$ kpc and the quasar flux ($F_{\rm Quasar}$) 
by integrating the PSF.
These are corrected for the Galactic extinction taken from \citet{schlegel98}.
Those fitting results whose estimated errors in galaxy flux exceed 2.5 mag are discarded; they are at the tail of the error distribution,
as most (about 70 \% in the $u$ band and more than 95 \% in the $g$, $r$, $i$, $z$ bands) of the objects with successful fits have
galaxy flux errors of less than 0.5 mag.


\begin{figure}
\epsscale{1.1}
\plotone{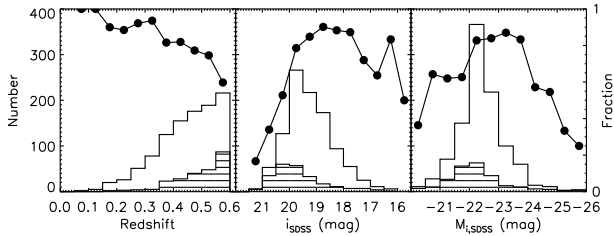}
\caption{Number of quasars in the initial sample as a function of redshift (left), $i_{\rm SDSS}$ (middle),
and $M_{i, {\rm SDSS}}$ (right) are shown by the histograms.
The hatched histograms represent the objects that fail to meet the good-fit criterion.
The good-fit fractions are shown by the dots connected by the solid lines (the scale is shown on the right axis).}
\label{fig:stat}
\end{figure}

For robust estimates of the galaxy $k$-correction and (stellar) mass-to-light ratio, we need to know 
the spectral energy distribution of each galaxy reasonably well.
To this end, we require that the fitting results be available in at least three of the $g$, $r$, $i$, and $z$ bands
(this is called the ``good-fit criterion" hereafter).
We decided not to consider the $u$-band results for further analysis, since the galaxy component is not visible 
in many objects in this band due to low signal-to-noise ratio and high quasar-to-galaxy contrast. 
Figure \ref{fig:stat} shows the number statistics of the initial sample and the fraction of the good-fit objects
as a function of redshift and magnitude.
The good-fit fraction decreases toward higher redshift and fainter magnitude, as expected, dropping to 
nearly 50\% at $z = 0.6$, $i_{\rm SDSS} = 20$ mag, or $M_{i, {\rm SDSS}} = -20$ mag.
The most luminous objects also tend to fail in the fitting because of the highest quasar-to-galaxy contrasts, 
but they are only a small fraction of the whole sample.
In total, 802 quasars or 77 \% of the initial sample meet the good-fit criterion.
For these objects, we estimate the absolute magnitudes and mass-to-luminosity ratios of the host galaxies
using the public $k$-correction code \texttt{kcorrect}, version 4.2 \citep{blanton07}.
Throughout this paper we express the absolute magnitudes of the host galaxies in the SDSS $g$ and $i$ bands 
shifted to $z = 0.3$, which are denoted as $^{0.3}M_g$ and $^{0.3}M_i$.
The color in these shifted bandpasses is $^{0.3}(g - i)$ $\equiv$ $^{0.3}M_g$ $-$ $^{0.3}M_i$.
Roughly speaking, the bandpasses of $^{0.3}M_g$ and $^{0.3}M_i$ are close to those of the original SDSS $u$ and $r$ bands, respectively.
The median conversion factor from $^{0.3}M_i$ to stellar mass $M_{\rm star}$ is found to be log $(M_{\rm star}/M_\odot) + 0.4\ ^{0.3}M_i = 2.09$.
Note that quasar absolute magnitudes are always $k$-corrected to $z = 0$ in this work.

\subsection{Systematic Errors \label{subsec:simulation}}

Here we estimate systematic errors arising from the above procedure with simulated quasar images.
We first select about 1300 galaxies on Stripe 82 from the SDSS DR10 \citep{ahn13} database, in such a way that they are distributed
fairly uniformly in the space of redshift and $g - i$ color in the range $0.1 < z < 0.6$ and $0 < g - i < 4$.
Although most of the DR10 galaxies with spectroscopic redshifts at $z > 0.3$ are red, with passive stellar population, we were able to
find a sufficient number of blue galaxies that cover the redshift and color distributions of the real quasar hosts (see below).
The galaxies (with SDSS model magnitudes $m_{\rm gal, orig}$) are extracted from the same imaging data as used in the above analysis, 
and their flux scales are
reduced so that the resultant galaxy magnitudes become $m_{\rm gal} = m_{\rm gal, orig} + {\Delta}m$ where ${\Delta}m = 0, 1, 2, 3, 4$ mag.
Since this procedure also reduces background noise, we add additional random Gaussian noise that restores the amplitude of the original 
background fluctuation.
Next we randomly pick a bright star from the SDSS Stripe 82 standard star catalog \citep{ivezic07} for each of the above galaxies, 
and add its image to the dimmed galaxy image by aligning the source centroids.
In order to reproduce a range of quasar-to-galaxy contrast, the stellar flux is reduced to 
$m_{\rm star} = m_{\rm gal} + {\Delta}m_{\rm contrast} + {\Delta}m_{\rm random}$ where ${\Delta}m_{\rm contrast} = -3, -2, -1, 0, 1, 2$ mag
and ${\Delta}m_{\rm random}$ is a random variable between $-0.5$ and $+0.5$ mag.
These superimposed galaxy $+$ star images comprise our synthesized quasar images, which are processed through the same
analysis algorithm as used for the real quasar sample.
For simplicity  in the rest of this subsection, we denote the input and output galaxy magnitudes by $m_0$ ($= m_{\rm gal}$ above) and $m$, 
respectively, and the input and output quasar magnitudes by  $m_0^{\rm QSO}$ ($= m_{\rm star}$ above) and $m^{\rm QSO}$, respectively.

\begin{figure}
\epsscale{1.0}
\plotone{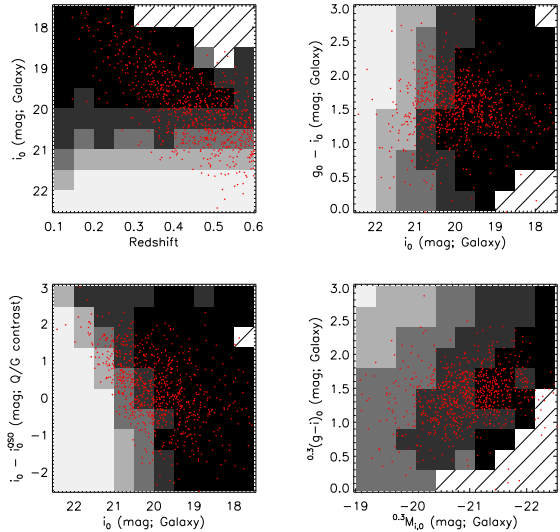}
\caption{Success rate of the fitting procedure as a function of redshift and magnitude $i_0$ (top left), 
magnitude $i_0$ and color $g_0 - i_0$ (top right), magnitude $i_0$ and quasar-to-galaxy contrast $i_0 - i_0^{\rm QSO}$ (bottom left), and
absolute magnitude $^{0.3}M_{i,0}$ and color $^{0.3}(g - i)_0$ (bottom right).
The rate decreases from $> 0.8$ in the black cells to $< 0.2$ in the lightest grey cells, while 
the hatched area shows where there are less than ten simulated galaxies in a cell.
The red dots represent our real quasars with successful fits.
}
\label{fig:simulation1}
\end{figure}

\begin{figure}
\epsscale{1.0}
\plotone{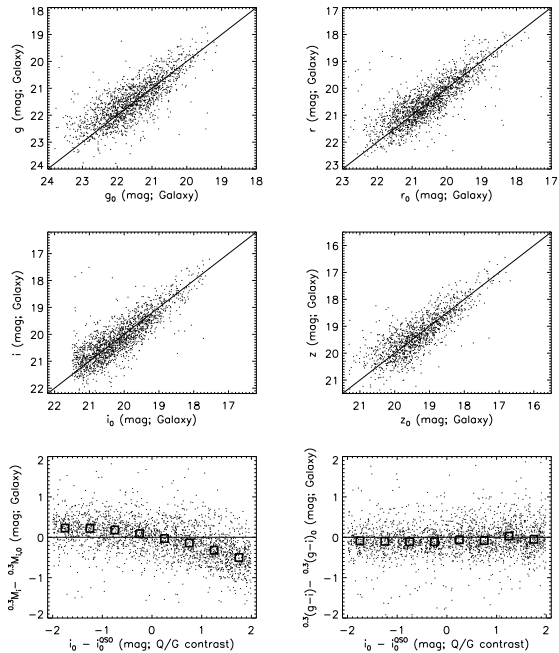}
\caption{Top and middle panels: comparisons of the input and output galaxy magnitudes in the $g$ (top left),
$r$ (top right), $i$ (middle left), and $z$ (middle right) bands from our simulations of galaxy-star superpositions.
The truncation of the $i_0$ distribution at 21.5 mag is caused by the selection criteria of the simulated quasars
used for these comparisons (see text).
Bottom panels: differences between the input and output values of $^{0.3}M_i$ (bottom left) and $^{0.3}(g-i)$ (bottom right) 
as a function of the quasar-to-galaxy contrast $i_0 - i_0^{\rm QSO}$ (dots).
The squares represent median values in bins.
In all the panels, the solid lines represent the locus where the input and output values are equal.
}
\label{fig:simulation2}
\end{figure}

Figure \ref{fig:simulation1} shows the success rate of the fitting procedure for the simulated quasars (the fraction of the simulated quasars meeting 
the good-fit criterion defined in the previous subsection) on a few projection planes of the parameter space.
The distribution of real quasars is superposed.
The success rate is most strongly dependent on galaxy magnitude; the fitting tends to fail for galaxies fainter than $i_0 \sim 21$ mag.
On the other hand, it is less dependent on redshift or galaxy color $g_0 - i_0$ at fixed galaxy magnitude.
For faint galaxies with $i_0 > 20$ mag, the success rate is higher at higher quasar-to-galaxy contrast, because the brighter quasar light
helps to pass the signal-to-noise ratio cut in the analysis algorithm.
Finally, on the galaxy CMD of $^{0.3}M_{i,0}$ and $^{0.3}(g - i)_0$, the success rate decreases toward the lower-luminous
and redder region where galaxies have the lowest apparent brightness in the optical bands.
We suspect that the real quasars that fail to meet the good-fit criterion, which comprise $\sim$23 \% of the initial sample, fall in those particular 
regions of the parameter space with low success rates.
The fraction of real quasars with successful fits drops in the transition area between the high and low success rates of the simulation, 
suggesting that the present simulation reproduces the behavior of the real quasar hosts reasonably well.

The next question is how accurately the host galaxy brightness is measured for those real quasars with successful fits.
We extract the simulated quasars in the parameter subspace where most of the real quasars with successful fits are found, 
i.e., $0.2 < z < 0.6$, $17.5 < i_0 < 21.5$, $0.5 < g_0 - i_0 < 2.5$, and $-2.0 < i_0 - i_0^{\rm QSO} < 2.0$, 
and compare the input and output magnitudes ($m_0$ and $m$) of the host galaxies.
The results are shown in Figure \ref{fig:simulation2}.
The two sets of magnitudes are fairly consistent with each other, with systematic offsets and rms scatters measured
to be ${\langle}g - g_0{\rangle} = -0.08\pm0.63$, ${\langle}r - r_0{\rangle} = -0.02\pm0.55$, ${\langle}i - i_0{\rangle} = +0.03\pm0.49$, 
and ${\langle}z - z_0{\rangle} = +0.01\pm0.49$.
The derived absolute magnitudes also show slight systematic offsets from their correct values, which we found are weakly correlated with 
quasar-to-galaxy contrast (shown in the bottom left panel of Figure  \ref{fig:simulation2}).
This is presumably caused by two competing effects; some of the galaxy light close to the nuclei can be attributed
wrongly to the quasar component (making $^{0.3}M_i - ^{0.3}M_{i,0}$ positive), while quasar light contaminates into host galaxies more severely
when the quasar-to-galaxy contrast is higher (making $^{0.3}M_i - ^{0.3}M_{i,0}$ negative).
However, this systematic error is small compared to the accuracy on which our conclusions stand, and we will see that
correcting for it as a function of the quasar-to-galaxy contrast has little effect on our conclusions.
Similarly, the systematic error of the derived absolute color ($^{0.3}(g-i) - ^{0.3}(g-i)_0 = -0.07$ mag in median) is too small to affect any of
our conclusions.

\section{Results \label{sec:results}}

\begin{figure}
\epsscale{1.0}
\plotone{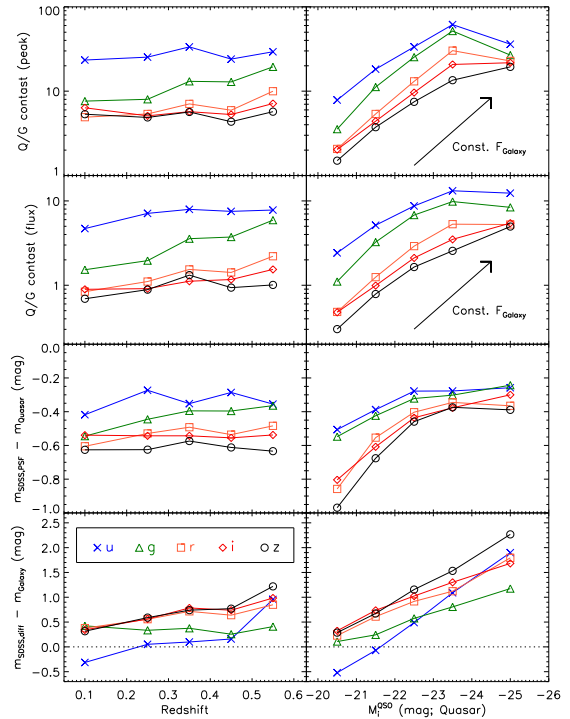}
\caption{Quasar-to-galaxy contrasts and associated magnitude differences as a function of redshift and quasar absolute magnitude.
The solid lines connected by crosses, triangles, squares, diamonds, and circles represent 
the $u$, $g$, $r$, $i$, and $z$ bands, respectively. 
Top: median peak intensity ratios at the source center ($R < 1$ pixel).
Second to top: median integrated flux ratios ($F_{\rm Quasar}$/$F_{\rm Galaxy}$).
The arrows represent the vector of constant $F_{\rm Galaxy}$.
Third to top: difference between the SDSS PSF magnitudes ($m_{\rm SDSS, PSF}$) and the decomposed 
quasar magnitudes ($m_{\rm Quasar}$).
Bottom: difference between $m_{\rm SDSS,diff}$ (corresponding to the SDSS model flux minus the SDSS PSF flux)
and the decomposed galaxy magnitudes ($m_{\rm Galaxy}$).}
\label{fig:qgcontrast}
\end{figure}

We show the quasar-to-galaxy contrasts and associated magnitude differences as a function of redshift and
quasar luminosity for the real quasars in Figure \ref{fig:qgcontrast}.
Here all the successful fitting results in each band are included, regardless of the good-fit criterion.
The median values of the contrasts (quasar brightness relative to galaxy brightness) in the ($u$, $g$, $r$, $i$, $z$) bands 
are (28.4, 14.2, 7.3, 5.8, 5.1) in peak intensity ratios and (7.5, 4.0, 1.6, 1.2, 1.0) in integrated flux ratios, respectively.
As expected, the contrast is significantly higher in the bluer bands.
While its correlation with redshift is relatively weak, that with the quasar luminosity (corresponding to $F_{\rm Quasar}$)
is remarkably strong. 
The slopes at $M_i < -24$ mag are close to the vector of constant $F_{\rm Galaxy}$, meaning that $F_{\rm Quasar}$
and $F_{\rm Galaxy}$ are primarily independent of each other.
\citet{hao05}  found a similar lack of correlation between nuclear and host luminosities for Seyfert galaxies at low redshifts ($z < 0.15$).
In our sample, the correlations flatten out at the higher luminosity where the fitting starts to fail for the highest-contrast objects.
Indeed, the actual contrasts would presumably be higher and bluer than estimated here if we had proper measurements of those 
highest-contrast objects.
Although the SDSS PSF magnitudes are commonly used to quantify the quasar brightness, they are found to be (0.33, 0.39, 0.51, 0.54, 0.61) 
mag brighter in median than the decomposed quasar magnitudes in the ($u$, $g$, $r$, $i$, $z$) bands.
The systematic difference is larger for lower-luminosity quasars.
On the other hand, the host brightness is sometimes approximated by the SDSS model flux minus the SDSS PSF flux.
We convert this quantity into magnitude $m_{\rm SDSS,diff}$, which is found to be (0.28, 0.33, 0.68, 0.82, 0.81) mag 
fainter than the decomposed galaxy magnitudes in median in the ($u$, $g$, $r$, $i$, $z$) bands. 
The discrepancy increases as the quasar luminosity increases.





\begin{figure*}
\epsscale{0.8}
\plotone{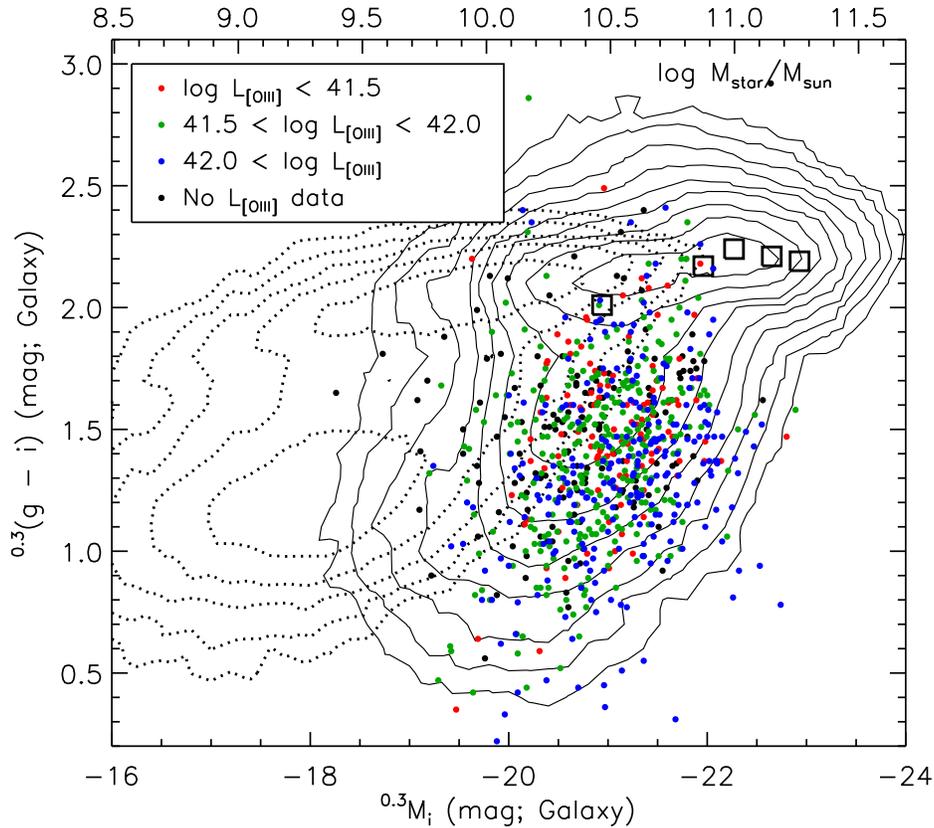}
\caption{Color--magnitude diagram of the quasar hosts (dots) compared to normal galaxies (contours).
The red, green, and blue dots correspond to the [\ion{O}{3}] luminosity of $L_{\rm [OIII]} < 10^{41.5}$ erg s$^{-1}$, 
$10^{41.5}$ erg s$^{-1}$ $< L_{\rm [OIII]} < 10^{42.0}$ erg s$^{-1}$, and $L_{\rm [OIII]} > 10^{42.0}$ erg s$^{-1}$, respectively.
The black dots represent objects with no $L_{\rm [OIII]}$ data.
The dotted and solid contours show the distributions of the SDSS normal galaxies (see text) at $z < 0.05$ and $0.1 < z < 0.6$, respectively.
They are drawn at logarithmically stepped levels of number density.
The squares represent the median $^{0.3}M_i$ and $^{0.3}(g - i)$ values of the normal galaxies at $z$ = 0.1--0.2,
0.2--0.3, 0.3--0.4, 0.4--0.5, and 0.5--0.6 from left to right.
The approximate stellar mass (log $M_{\rm star}$/$M_\odot$) is shown on the upper axis.
}
\label{fig:cmd}
\end{figure*}

\begin{figure*}
\epsscale{0.8}
\plotone{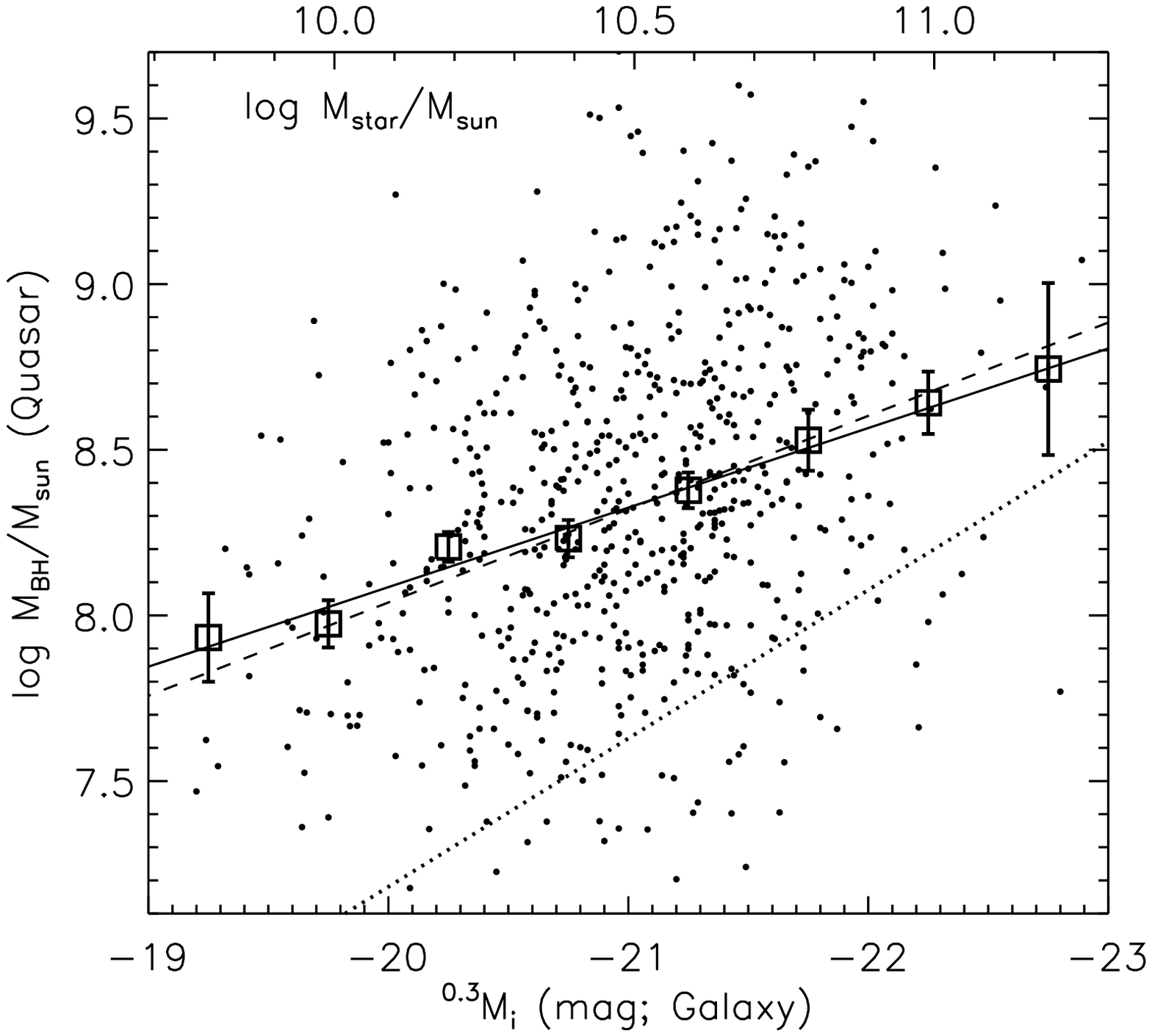}
\caption{Relation between the mass of SMBHs and the host absolute magnitude $^{0.3}M_i$ of the quasars (black dots).
The squares and the solid line represent the mean values and the best-fit linear relation, respectively.
The dashed line shows the best-fit relation obtained when $^{0.3}M_i$ is corrected for its systematic error
as a function of the measured quasar-to-galaxy contrast; this correction is very minor.
The dotted line shows the local relation for quiescent galaxies \citep{haring04}.
The approximate stellar mass (log $M_{\rm star}$/$M_\odot$) is shown on the upper axis.
}
\label{fig:magorrian}
\end{figure*}

Figure \ref{fig:cmd} shows the CMD of the host galaxies.
The approximate stellar mass is shown on the upper axis, using the conversion described at the end of Section \ref{subsec:imageanalysis}.
The distribution of normal galaxies from the SDSS spectroscopic sample at $0.1 < z < 0.6$, taken from the MPA-JHU DR7 
catalog\footnote{http://www.mpa-garching.mpg.de/SDSS/}, is also shown for reference.
We note that their redshift distribution is not well matched to that of the quasar host galaxies; about 80 \% of the reference sample 
is at $z < 0.3$ where very few of the quasar hosts are found.
If we limit to $z > 0.3$, there is not a large and complete enough sample of normal galaxies in SDSS to serve as a good reference in this CMD.
The median $^{0.3}M_i$ and $^{0.3}(g - i)$ values of the reference sample at each redshift are plotted in the figure.
We also plot the MPA-JHU galaxies at lower redshifts ($z < 0.05$) in order to show the location of the blue cloud in the local universe
(the blue cloud is not clear at higher redshifts due to the limited depth of SDSS).
The hosts of our quasars occupy a unique location in this diagram;
they are almost exclusively massive with $10^{10} M_{\odot} < M_{\rm star} < 10^{11} M_{\odot}$,
and have colors ranging from the blue cloud to below the bottom of the red sequence.
Compared to the majority of the normal galaxies with similar stellar mass, those quasar hosts are forming stars much more efficiently.
Although objects with different quasar power (measured by the [\ion{O}{3}] $\lambda$5007 luminosity $L_{\rm [O III]}$) are well 
mixed with each other, there seems to be a slight offset between the higher- and lower-$L_{\rm [O III]}$ objects.
We will return to this issue later.


With the decomposed galaxy components in hand, we can also check the relation between the mass of SMBHs and the host luminosity (stellar mass).
They are plotted in Figure \ref{fig:magorrian}.
Although the data points show significant scatter, there is a clear positive correlation between the two quantities.
Thus we have recovered the $M_{\rm BH} - M_{\rm star}$ relation in optically luminous quasars.
The best-fit linear relation is
\begin{equation}
{\rm log}\ \frac{M_{\rm BH}}{10^8 M_\odot} = 0.09 - 0.24\ (^{0.3}M_i + 20) ,
\end{equation}
or
\begin{equation}
{\rm log}\ \frac{M_{\rm BH}}{10^8 M_\odot}  = 0.03 + 0.60\ {\rm log}\ \frac{M_{\rm star}}{10^{10} M_\odot}  .
\end{equation}
Compared to the local $M_{\rm BH} - M_{\rm star}$ relation of quiescent galaxies \citep{haring04}, those quasar hosts are found to be
under-massive for a given SMBH mass.
Considering that our fitting procedure tends to fail for less-luminous host galaxies, this trend 
may be even stronger than appears in Figure \ref{fig:magorrian}.
As shown in the figure, this result hardly changes when we correct for the systematic errors of $^{0.3}M_i$ as a function of the measured
quasar-to-galaxy contrast (see Section \ref{subsec:simulation}).
However, the relation measured here could be subject to serious selection effects and other observational biases, as discussed below.


\section{Discussion \label{sec:discuss}}

\subsection{Quasar Light in Host Galaxies \label{subsec:quasarlightinhost}}

\begin{figure}
\epsscale{1.0}
\plotone{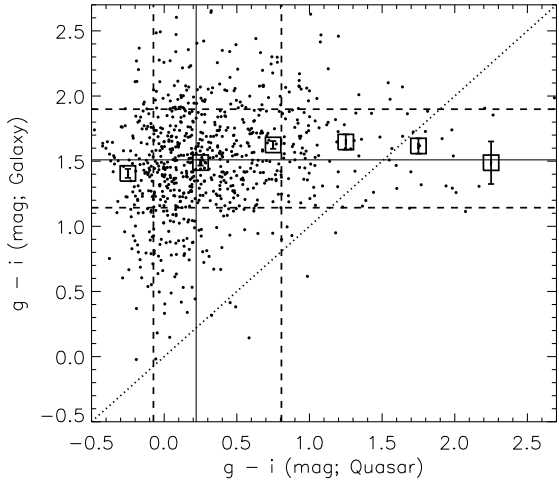}
\caption{$g - i$ colors of the quasars (the PSF components) vs. those of the host galaxies (the S\'{e}rsic components).
All the objects with successful fits in the $g$ and $i$ bands are included.
The squares represent the mean values.
The vertical and horizontal lines show the distribution of the quasar and galaxy colors, respectively; 68.3 \% of the data points
around the median value (solid line) are distributed between the two dashed lines.
The dotted line shows the locus where the colors of the two components are identical.}
\label{fig:gmicor}
\end{figure}

The simulations we carried out in Section \ref{subsec:simulation} show that the measured host galaxy magnitudes are only slightly biased.
In this subsection we explore this further, by looking for trends in the observed data that might indicate contamination of
the galaxy light from the quasar.
There are two possible causes of such a contamination.
One is imperfect nucleus/host decomposition of the source profile, due to the limited precision of the PSF estimation
and/or the fact that the galaxy may not be perfectly fit by a S\'{e}rsic profile.
Our fits may tend to overestimate the S\'{e}rsic component since the S\'{e}rsic, with three parameters, has more flexibility to 
fit a given profile than does the PSF.
On the other hand, if there is a compact stellar core at the galactic center on top of the S\'{e}rsic profile, such as
a nuclear starburst, then it would be included in the PSF and lead to underestimation of the galaxy component.
These two effects bias the measured host brightness in opposite directions.
The other possible cause of the contamination is scattering of quasar light by the interstellar medium of the host galaxy.
Unlike the first effect, this is an astrophysical effect and is an interesting subject in itself.
Contamination of quasar light in the photometry of host galaxies, originating from either of these effects, would 
reveal itself most clearly in colors, since quasars are much bluer than galaxies in general.
However, as we find in Figure \ref{fig:gmicor}, the color distributions of the decomposed quasars (PSF components) and 
host galaxies (S\'{e}rsic components) seem to be uncorrelated.
The variation of the mean galaxy colors is only about $\pm0.1$ mag as the quasar colors change by 3 mag.
Note that the slight positive correlation of the quasar and galaxy colors is expected, since our fitting analysis tends to fail
for the bluest quasars with the reddest hosts due to high quasar-to-galaxy contrast in the $g$ band.
Furthermore, we have already seen that the magnitudes and colors of galaxies hosting the quasars with a range of 
[\ion{O}{3}] luminosity are relatively well mixed (see Fig \ref{fig:cmd}).
These facts imply that quasar light contamination is not a large effect.

\begin{figure}
\epsscale{1.0}
\plotone{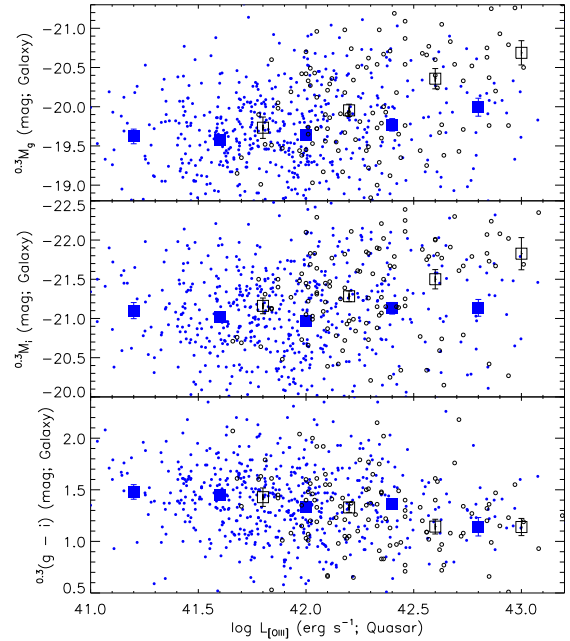}
\caption{Absolute magnitudes $M^{0.3}_g$ (top), $M^{0.3}_i$ (middle) and colors $^{0.3}(g - i)$ (bottom) 
of quasar host galaxies as a function of [\ion{O}{3}] $\lambda$5007 luminosity.
The dots and open circles represent the unobscured (type-1) quasars of this work and SDSS type-2 quasars \citep{reyes08} at $0.3 < z < 0.6$, 
respectively.
The filled and open squares represent the mean values of the type-1 and type-2 samples, respectively.}
\label{fig:fscattered_type12}
\end{figure}

A straightforward way to check the effect of the imperfect nucleus/host decomposition is to compare the measured host properties
of quasars with and without measurable nuclear contribution. 
To this end, we compare the host magnitudes and colors of our sample to those of SDSS type-2 (obscured) quasars \citep{reyes08} as shown 
in Figure \ref{fig:fscattered_type12}.
The redshifts of both samples are limited to $0.3 < z < 0.6$ (where most of our sample is located) for this comparison.
The lack of type-2 quasars at low [\ion{O}{3}] luminosity is due to the luminosity threshold imposed by Reyes et al. 
($L_{\rm [O III]} \ga 10^{41.9}$ erg s$^{-1}$).
It is seen that the type-2 quasars are up to 0.5 mag more luminous than the hosts of type-1 objects.
Since this excess brightness is higher at higher quasar luminosity, it may be due to quasar light scattered at the central part of the galaxy 
or nuclear starburst associated with AGN, which are included in the PSF component for the type-1 objects.
If it has a stellar origin, then the magnitudes (stellar mass) of our quasar hosts could be under-estimated by $< 0.5$ mag ($< 0.2$ dex).
However, such a small reduction in host stellar mass doesn't change any of our conclusions significantly.
On the other hand, the colors of the two types of quasars are almost perfectly consistent with each other, suggesting that our approach 
is very efficient in eliminating the nuclear contribution to host colors.
The weak anti-correlation between the colors and [\ion{O}{3}] luminosity can be explained by either AGN-associated
star formation or scattered quasar light.

Currently the most reliable way to quantify the scattered quasar light is via polarimetric measurements.
Based on the optical broadband- and spectro-polarimetric observations of SDSS type-2 quasars, 
\citet{zakamska05, zakamska06} confirmed the presence of significant amounts of scattered quasar light in host galaxies,
and suggested a typical scattering efficiency of $\epsilon_{\rm sca} \sim 1$ \%.
As a trial we re-calculated the absolute magnitudes and colors of our quasar hosts assuming $\epsilon_{\rm sca} = 1$ \% 
(i.e., we subtracted 1 \% of the decomposed PSF flux from the S\'{e}rsic flux before the $k$-correction),
and found that the median values of $^{0.3}M_i$ and $^{0.3}(g - i)$ for the hosts change only by $+0.01$ and $+0.02$ mag, respectively.
This is not surprising indeed, since the median quasar-to-galaxy contrasts of integrated flux
are at most 4 in the $g$, $r$, $i$, and $z$ bands;
only 1 \% of  the quasar brightness cannot make a significant impact on the host brightness.
In the extreme case of $\epsilon_{\rm sca} = 10$ \%, $^{0.3}M_i$ and $^{0.3}(g - i)$ would become
larger than their original values by $+$0.10 mag and $+$0.21 mag, respectively.
However, no compelling evidence for such a high scattering efficiency has been reported to date.
Hence we conclude that the scattered quasar light has no significant impact on our results.

\subsection{Massive Star-Forming Host Galaxies}

Figure \ref{fig:cmd} demonstrates that the quasars of our sample are preferentially hosted by massive ($M_{\rm star} > 10^{10} M_\odot$) 
galaxies with active star formation.
Massive hosts have also been found for less luminous, narrow-line AGNs in the SDSS spectroscopic sample
at lower redshifts \citep[$z < 0.3$;][]{kauffmann03}.
We note that our fitting analysis fails for about 23 \% of the initial sample, most of which are too faint to 
give sufficient signal-to-noise ratios in the combined quasar and galaxy profiles.
These faint sources could have systematically different host properties from those plotted in Figure \ref{fig:cmd}.
Nevertheless, we conclude that massive star-forming hosts are a common property of the majority, if not all, of low-$z$ ($z < 0.6$) SDSS quasars.
At lower luminosity ($M_u (z = 0.05) > -22$ mag) and lower redshifts ($z < 0.11$), \citet{trump13} reached a very similar conclusion to ours.
Their method of AGN/host decomposition depends critically on the assumption that the AGN hosts have the same morphological
structure as inactive galaxies on average, while our approach fit for the surface brightness profile of the hosts directly.

It is obvious that our conclusions do not apply to any quasar populations that are not included in the SDSS sample.
However, the completeness of the SDSS selection for unobscured quasars is estimated to be fairly high; \citet{vandenberk05} reported 
an overall completeness of about 89 \% or higher to a limiting magnitude of $i = 19.1$ mag
\citep[corresponding to the limiting magnitude of the SDSS-I selection for low-$z$ quasars; see also][]{richards06}.
Furthermore, Stripe 82 has been observed for quasars more completely than any other area of the SDSS coverage 
using a variety of variability and color selection algorithms \citep{ross12,palanque11,palanque13}.
Hence our conclusions may be generalized to all those unobscured quasars whose luminosity and redshifts are covered by the present sample.

Previous studies of narrow-line AGNs at low redshifts \citep[$z < 0.3$, e.g.,][]{kauffmann03,salim07,schawinski10}
have already found that the fraction of AGNs is higher in the blue cloud and the green valley than in the red sequence on the CMD.
\citet{kauffmann03} found that the hosts of their low-luminosity type-2 AGNs have similar stellar content to normal early-type
galaxies, while those of the high-luminosity AGNs have much younger stellar populations.
\citet{salim07} reported that the hosts of the most powerful AGNs in their sample have comparable stellar mass and star formation
rates (SFRs) to galaxies at the massive end of the blue cloud, while those of the weaker AGNs have considerably lower SFRs,
extending into the red sequence.
Our finding of almost no quasar hosts on the red sequence is consistent with these results, as the quasars are 
an-order-of-magnitude higher-luminosity (measured by $L_{\rm [O III]}$) counterparts to the narrow-line AGNs studied by the above authors.
The bluer hosts of more active AGNs might point to the presence of AGN-induced star formation predicted in numerical simulations 
\citep{wagner12,wagner13}.


On the other hand, studies of optically faint X-ray AGNs at redshifts similar to or larger than those of our sample 
(up to $z \sim 1$) have found that their hosts 
are preferentially in the green valley and the red sequence \citep[e.g., ][]{nandra07, georgakakis08, silverman08}.
This may suggest that different AGN types are hosted by different types of galaxies.
If the optically luminous quasars and X-ray AGNs are on a single evolutionary path of hosts transiting from 
the blue cloud to the red sequence, then our result might suggest that optically luminous quasars precede 
X-ray AGNs in an evolutionary sequence \citep[see also][]{shen07,hickox09,goulding13}.
According to the merger-driven co-evolution scenario of galaxies and SMBHs \citep{hopkins06},
interactions/mergers of gas-rich galaxies first activate star formation and AGN, and then produce a dust-obscured, IR-luminous phase
sometimes observed as ULIRGs.
The narrow-line AGN phase may then follow.
The systems further go into an optically luminous quasar phase when the dust is expelled by radiation pressure and the active nuclei become visible, 
and finally migrate into the red sequence as quasar and star-forming activity fades away.
The blue host colors we found support the idea that quasars coincide with active star formation in host galaxies.
Although the quasar-driven quenching of star formation (negative AGN feedback) has been proposed by many authors, 
such a process never becomes apparent in stellar properties during the most optically-luminous (i.e., quasar) phase of AGNs.

 

Some authors suggest that the apparent clustering of AGN hosts in the green valley and the red sequence is due to selection effects; 
if one works with a stellar mass-limited sample instead of a flux-limited sample, then the AGN fraction in the blue cloud would become 
comparably high to other regions of the CMD \citep{silverman09, xue10}.
In addition, \citet{cardamone10} argue that most of the green AGN hosts have colors consistent with dusty star-forming galaxies and
their intrinsic (dust-corrected) colors are located in the blue cloud.
These studies point to a larger fraction of blue hosts than previously thought for obscured AGNs.
If the above effects are also present in our analysis, then it would further strengthen our conclusion that quasar hosts are
bluer and forming stars more efficiently than are the majority of normal galaxies.

\begin{figure}
\epsscale{1.0}
\plotone{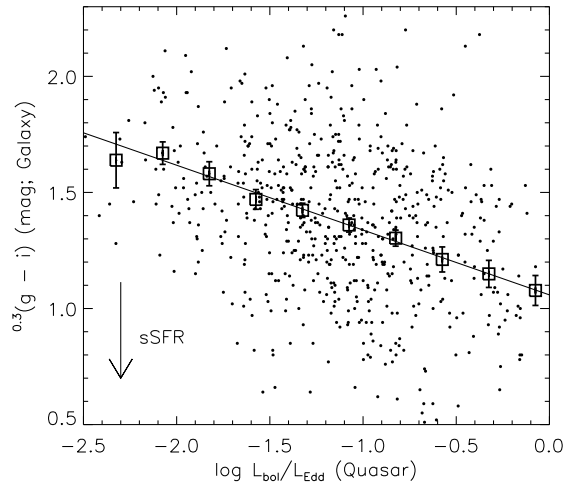}
\caption{Host color vs. Eddington ratio of the quasars (dots).
The mean values and the best-fit linear relation are represented by the squares and the solid line, respectively.
Specific SFR increases toward the bottom of this plot.}
\label{fig:ReddvsSFR}
\end{figure}

In Figure \ref{fig:ReddvsSFR}, we plot the host colors versus the quasar Eddington ratios ($R_{\rm Edd}$) in order to further investigate 
the possible interrelation between nuclear and star-forming activity.
Very roughly speaking, the vertical axis corresponds to $-2.5$ times the logarithm of specific SFR 
($-2.5$ log [SFR/${M_{\rm star}}$]) with a constant offset
since log SFR $\approx$ log (UV luminosity) $= -0.4\ ^{0.3}M_g$ $+$ constant and log $M_{\rm star} \approx$ log (red optical/NIR luminosity) 
$= -0.4\ ^{0.3}M_i$ $+$ constant. 
If one assumes that the $M_{\rm BH} - M_{\rm star}$ relation holds at every epoch of galaxy evolution (i.e., $M_{\rm BH} (t) \propto M_{\rm star}$ (t) at
every time $t$), then
\begin{equation}
{\rm sSFR} (t) = \frac{\dot{M}_{\rm star} (t)}{M_{\rm star} (t)} = \frac{\dot{M}_{\rm BH} (t)}{M_{\rm BH} (t)} \propto R_{\rm Edd} (t) .
\label{eq:ReddvsSFR}
\end{equation}
The Eddington ratio and sSFR do show a positive correlation in Figure \ref{fig:ReddvsSFR}, but the slope is much shallower than expected from 
Equation (\ref{eq:ReddvsSFR}); sSFR only doubles ($^{0.3}(g - i)$ increases by about 0.7) as $R_{\rm Edd}$ increases by more than a factor of 100.
This implies that the rise and fall of AGN and star formation are not precisely synchronized, although the assembly of SMBH and host stars 
keep pace with each other when averaged over time.

\subsection{$M_{\rm BH} - M_{\rm star}$ Relation}

\begin{figure}
\epsscale{1.0}
\plotone{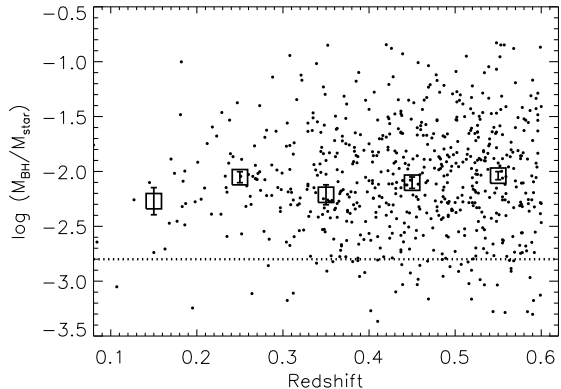}
\caption{$M_{\rm BH}/M_{\rm star}$ ratio as a function of redshift (dots).
The squares represent the mean of the sample.
The dotted line represents the local relation of quiescent galaxies \citep{haring04}.}
\label{fig:magorrian_z}
\end{figure}

Figure \ref{fig:magorrian} demonstrates that the hosts of the SDSS quasars are under-massive for a given SMBH mass compared to the local relation.
The discrepancy becomes larger at smaller mass, resulting in a shallow slope of the $M_{\rm BH} - M_{\rm star}$ relation.
Such a systematic difference between the local relation and the relation we found could originate from various effects.
The first is cosmological evolution during the time span of 5.7 Gyr between $z = 0.6$ and $z = 0.0$.
However, we found little redshift evolution of the $M_{\rm BH}/M_{\rm star}$ ratio in the range probed by our sample ($0.1 \la z < 0.6$)
as shown in Figure \ref{fig:magorrian_z}.
The ratio is offset from the local relation by the same amount even at $z < 0.3$, which casts doubt on the interpretation that
the offset is due entirely to evolutionary effects.
The second is an intrinsic difference between active and quiescent galaxies; although the $M_{\rm BH}$ estimators of active SMBHs are calibrated so
that the local AGNs fall on the $M_{\rm BH} - M_{\rm star}$ relation of local quiescent galaxies, it is not clear whether the calibration remains robust
at the higher-redshift universe.
The third is selection effects and observational biases, as we now discuss.


Previous studies of the relation between $M_{\rm BH}$ and $M_{\rm star}$ (or the central stellar dispersion $\sigma_*$ or the spheroid mass $M_{\rm sph}$) 
beyond the local universe are still very controversial.
Many observations at $z < 1$ \citep[e.g., ][]{treu04, woo06,woo08,canalizo12} and beyond \citep[e.g., ][]{peng06,shields06,bennert11}
even up to $z \sim 6$ \citep{walter04,wang13} suggest that the $M_{\rm BH}$/$M_{\rm sph}$ ratio
increases toward high redshift.
Our results are qualitatively consistent with these studies, although our $M_{\rm star}$ includes not only a spheroid but also a disk component if it exists.
Attempts have been made to quantify the evolution of the ratio in the form of $M_{\rm BH}$/$M_{\rm sph}$ $\propto (1 + z)^\gamma$, but $\gamma$
varies significantly from paper to paper; for example, $\gamma = 2.07\pm0.76$ \citep[measured at $0 < z < 2$;][]{mclure06}, $\gamma = 0.68\pm0.12^{+0.6}_{-0.3}$
\citep[$1 < z < 2.2$;][]{merloni10}, or $\gamma = 1.4\pm0.2$ \citep[$0 < z < 4.5$;][]{bennert10}.
If $M_{\rm BH}$/$M_{\rm sph}$ increases toward high redshift, this suggests that SMBHs grow, or are assembled, earlier than
their host galaxies.
However, other studies claim no such evolution. 
\citet{shields03} showed that the local $M_{\rm BH} - \sigma_*$ relation holds up to $z \sim 3$.
Using over 900 AGNs drawn from SDSS, \citet{shen08} also found no evolution of the $M_{\rm BH} - \sigma_*$ relation at $z < 0.4$.
Similar conclusions were reached for AGNs at $0.5 < z < 1.2$ by \citet{schramm13} and at $2 \la z \la 3$ by \citet{adelberger05}.


At least some of the above controversy may be attributed to various observational biases \citep[e.g.,][]{salviander07}.
In a flux-limited sample, quasar luminosity and $M_{\rm BH}$ are systematically larger in higher-redshift objects.
If the $M_{\rm BH} - M_{\rm star}$ relation is not well calibrated at high masses in the local universe, then 
the $M_{\rm BH}$/$M_{\rm star}$ ratio would deviate systematically at high redshifts.
If the single-epoch $M_{\rm BH}$ estimators are not accurately calibrated over a wide dynamic range of quasar luminosity, then it would also give
rise to a redshift-dependent bias.
\citet{shen10} demonstrated that the uncertainty in the single-epoch $M_{\rm BH}$ estimates 
could have a significant impact on the derived $M_{\rm BH} - M_{\rm star}$ relation.
Various quality cuts used to select the sample may also result in a redshift-dependent bias.
Furthermore, for a given galaxy mass, more massive SMBHs have higher chance of being selected
if they tend to have higher AGN luminosity \citep{lauer07}.
When coupled with a bottom-heavy galaxy mass function, this can produce a Malmquist-type bias and result in a higher $M_{\rm BH}$/$M_{\rm star}$ 
ratio at higher redshifts \citep[see also][]{shen13}.
A similar but different source of bias was pointed out by \citet{schulze11}, in which the decreasing fraction of active SMBHs at higher mass results in 
a larger fraction of lower-mass SMBHs being selected at fixed bulge properties.
They gave a comprehensive discussion of various selection biases and
demonstrated that, while the observed departures of the $M_{\rm BH}$/$M_{\rm bulge}$ ratio from the local relation are sometimes very large
at high redshifts, they could be all attributed to the accumulated effects of various selection biases.
The fundamental problem is that our current knowledge about SMBHs beyond the local universe is necessarily based only on AGNs above 
a certain observed flux.
We do not really know whether the points in Figure \ref{fig:magorrian} are representative of, or just a part of the envelope of, 
the underlying distribution of all the (active and inactive) SMBHs at the relevant redshifts.

\section{Summary and Conclusions\label{sec:summary}}

We have analyzed the photometric properties of galaxies hosting optically luminous, unobscured quasars at $z < 0.6$.
The sample was selected from the fifth SDSS quasar catalog \citep{schneider10} and the first BOSS quasar catalog \citep{paris12}
on Stripe 82, where deep co-added SDSS images are available.
Each quasar image is decomposed into nucleus and host galaxy using the PSF and the S\'{e}rsic models.
802 objects (77 \% of the initially selected sample) are successfully fitted with our approach in at least three of the $g$, $r$, $i$, and $z$ bands,
which constitute our final sample.
We confirmed that the effect of quasar light contamination due to the imperfect nucleus/host decomposition is not very large.
The systematic errors in the measured galaxy absolute magnitudes and colors are estimated to be less than 0.5 mag and 0.1 mag,
respectively, with simulated quasar images.
Scattered quasar light cannot affect host brightness as long as the scattering efficiency is about 1 \% or less.
Our main conclusions are as follows.
\begin{enumerate}
\item The SDSS quasars are almost exclusively hosted by massive galaxies with $M_{\rm star} > 10^{10} M_{\odot}$.
This is consistent with the results of previous studies on less luminous, narrow-line AGNs.
The quasar hosts are also very blue and almost absent on the red sequence, which is in stark contrast to the color--magnitude 
distribution of normal galaxies.
The fact that more active AGNs reside in bluer galaxies may suggest that negative AGN feedback, if it exists, is not concurrent with
the most optically luminous phase of AGN.
\item There is a positive correlation between $M_{\rm BH}$ and $M_{\rm star}$ of the SDSS quasars.
However, the $M_{\rm BH} - M_{\rm star}$ relation is offset toward larger $M_{\rm BH}$ or smaller $M_{\rm star}$ compared to the local relation.
While this could indicate that SMBHs are assembled or grow earlier than their host galaxies, we argue that it is not clear whether the sample is 
representative of the whole active and inactive SMBHs at the relevant redshifts due to various observational biases.
\end{enumerate}

The present work demonstrates that luminosity and colors of quasar host galaxies can be extracted from optical multi-band images 
with moderate spatial resolution.
We will expand the analysis into higher redshifts and lower luminosity with the wide-field imaging data obtained in 
the coming Subaru/Hyper Suprime-Cam (HSC) survey, which is scheduled to start in early 2014. 
The Wide layer of the HSC survey plans to observe 1400 deg$^2$ of the SDSS footprint with an expected depth of 
$i = 25.9$ mag (5$\sigma$, 2\arcsec\ aperture) and a PSF full width at half maximum of less than 0\arcsec.7.
Thus these data will significantly supersede SDSS both in depth and image quality.



\acknowledgments

We are grateful to Jenny Greene and Yue Shen for fruitful discussions and suggestions.
We thank an anonymous referee for his/her very useful comments.
This work was supported by Grants-in-Aid for Young Scientists (A) (22684995) and for Scientific Research on Innovative Areas (24111705) from the Japan Society for the Promotion of Science (JSPS).
YM was a JSPS research fellow during part of this work and was supported by a Grant-in-Aid for JSPS Fellows (SPD, 25-1646).

Funding for the SDSS and SDSS-II has been provided by the Alfred P. Sloan Foundation, the Participating Institutions, the National Science Foundation, the U.S. Department of Energy, the National Aeronautics and Space Administration, the Japanese Monbukagakusho, the Max Planck Society, and the Higher Education Funding Council for England. The SDSS Web site is http://www.sdss.org/.

The SDSS is managed by the Astrophysical Research Consortium for the Participating Institutions. The Participating Institutions are the American Museum of Natural History, Astrophysical Institute Potsdam, University of Basel, University of Cambridge, Case Western Reserve University, University of Chicago, Drexel University, Fermilab, the Institute for Advanced Study, the Japan Participation Group, Johns Hopkins University, the Joint Institute for Nuclear Astrophysics, the Kavli Institute for Particle Astrophysics and Cosmology, the Korean Scientist Group, the Chinese Academy of Sciences (LAMOST), Los Alamos National Laboratory, the Max-Planck-Institute for Astronomy (MPIA), the Max-Planck-Institute for Astrophysics (MPA), New Mexico State University, Ohio State University, University of Pittsburgh, University of Portsmouth, Princeton University, the United States Naval Observatory, and the University of Washington.

Funding for SDSS-III has been provided by the Alfred P. Sloan Foundation, the Participating Institutions, the National Science Foundation, and the U.S. Department of Energy Office of Science. The SDSS-III Web site is http://www.sdss3.org/.

SDSS-III is managed by the Astrophysical Research Consortium for the Participating Institutions of the SDSS-III Collaboration including the University of Arizona, the Brazilian Participation Group, Brookhaven National Laboratory, University of Cambridge, Carnegie Mellon University, University of Florida, the French Participation Group, the German Participation Group, Harvard University, the Instituto de Astrofisica de Canarias, the Michigan State/Notre Dame/JINA Participation Group, Johns Hopkins University, Lawrence Berkeley National Laboratory, Max Planck Institute for Astrophysics, Max Planck Institute for Extraterrestrial Physics, New Mexico State University, New York University, Ohio State University, Pennsylvania State University, University of Portsmouth, Princeton University, the Spanish Participation Group, University of Tokyo, University of Utah, Vanderbilt University, University of Virginia, University of Washington, and Yale University.

\end{document}